\documentclass{sig-alt-full}

\begin{document}

\conferenceinfo{WMuNeP'08,} {October 27, 2008, Vancouver, BC, Canada.} 

\CopyrightYear{2008}

\crdata{978-1-60558-238-2/08/10} 

\title{Maximally Radio-Disjoint Multipath Routing for Wireless
Multimedia Sensor Networks}

\numberofauthors{1} 
\author{
\alignauthor 
Moufida Maimour 
\affaddr{CRAN laboratory}\\
\affaddr{Nancy University, CNRS, France}\\
\email{Moufida.Maimour@cran.uhp-nancy.fr}
}

\maketitle
\begin{abstract}
In wireless sensor networks, bandwidth is one of precious resources to
multimedia applications. To get more bandwidth, multipath routing is
one appropriate solution provided that inter-path interferences are
minimized. In this paper, we address the problem of interfering paths
in the context of wireless multimedia sensor networks and consider
both intra-session as well as inter-session interferences. Our main
objective is to provide necessary bandwidth to multimedia applications
through non-interfering paths while increasing the network
lifetime. To do so, we adopt an incremental approach where for a given
session, only one path is built at once. Additional paths are built
when required, typically in case of congestion or bandwidth
shortage. Interference awareness and energy saving are achieved by
switching a subset of sensor nodes in a {\em passive state} in which
they do not take part in the routing process. Despite the routing
overhead introduced by the incremental approach we adopt, our
simulations show that this can be compensated by the overall
achieved throughput and the amount of consumed energy per correctly
received packet especially for relatively long sessions such as
multimedia ones. This is mainly due to the fact that a small number of
non-interfering paths allows for better performances than a large
number of interfering ones.
\end{abstract}


\noindent
{\bf General Terms:} Algorithms, Design, Performance

\section{Introduction}\label{intro}

Recent advances in micro-electronics and wireless communications in
addition to the availability of inexpensive CMOS cameras and
microphones, allowed for the emergence of wireless multimedia sensor
networks (WMSN)\cite{WMSNSurvey}. These latters can significantly
enhance a wide range of applications like object detection,
surveillance, recognition, localization, and tracking. Such
applications require a significant effort in developing new approaches
well adapted to both wireless sensor networks and multimedia transport
specific characteristics. On the one hand, sensor networks are very
constrained in terms of energy, processing power and bandwidth. On
the other hand, multimedia applications are resources hungry and are
very demanding of bandwidth. 

One solution to provide sufficient bandwidth to multimedia
applications in WSN and hence improve their quality, is multipath
routing that has been a hot research area for years. Most of multipath
routing protocols in wireless sensor \cite{Highly-DD}\cite{LMR} and
generally in ad hoc networks \cite{MR-DSR-based}\cite{AODV-BR} were
targeted to failure tolerance. Additional paths are maintained to
serve as backup on primary path failure. Few of them like SMR
\cite{SMR} consider the use of multiple paths concurrently. In SMR,
two or more maximally disjoint routes are used
simultaneously. Disjointness \cite{Yeming2006}\cite{SrinivasMobicom03}
allows for a more balanced traffic in the network however, it does not
deal with interferences. The selected paths may be highly interfering
if their respective links are too close. Consequently, the overall
achieved throughput is far from being the summation of the bandwidth
offered independently by the different paths \cite{BoutabaICC06}.

Because of the shared nature of the medium and the non centralized
random access protocols in ad hoc wireless network, interferences are
likely to appear when two active (transmitting/receiving) nodes are in
the radio range of each other. 
One approach to reduce interferences is the design of routing
protocols with new metrics that integrate interferences in their paths
cost. Some of these metrics are ETX (the expected transmission count
metric)\cite{LQSR} and ETT (the expected transmission time
metric)\cite{MR-LQSR}. Extensions of these metrics have recently been
proposed considering the availability of a multi-channel facility in
mesh networks \cite{MR-LQSR}\cite{CAM}. More recently
\cite{MoorsMeshTech2007} proposed WIM (Weighted Interference Multipath
metric) that allows for considering {\em spatial diversity} through
introducing the number of neighbors in the estimation of
interferences. 

All the previously cited works on interference-aware routing, rely on
metric estimations that require frequent periodic probing messages
exchange. This is not suitable to WSN characterized by their scarce
energy and limited processing capabilities. In fact, most of these
metrics are targeted to mesh networks where these limitations do not
hold. Moreover, they adopt a source-routing approach which is, once
again, not suitable to WSN because of RREQ packets size requiring more
energy to be transmitted/received. Finally, proposals relying on a
multi-channel facility are better suitable to mesh networks while they
are not adapted to WSN due to the overhead and energy consumption
needed for channel switching.

In this paper, we consider the problem of interfering paths in
WMSN. We assume that only one channel is available and thus we do not
address intra-path interferences. Instead, we focus on inter-path
interferences and consider both intra-session (for one source) and
inter-session (typically, multiple sources) interferences. Our main
objective is to provide necessary bandwidth to multimedia applications
through non-interfering (radio disjoint \cite{RadioDisjoint}) paths
while increasing the network lifetime. To achieve our twofold goal, we
chose to adopt an incremental approach where only one path is built at
once for a given session. Additional paths are built when required,
typically in case of path congestion or lack of bandwidth. When a
given path is selected to be used, all nodes interfering with it are
put in a {\em passive state}. Passive nodes do not further take part
in the routing process so they could not be used to form a new path
that consequently, will not interfere with previously built
ones. Moreover, passive nodes can be put in sleep or idle modes, thus
allowing for energy saving and hence increasing the network lifetime.

Despite the routing overhead introduced by the incremental approach we
adopted, our simulations showed that this can be compensated by the
overall achieved throughput and the amount of consumed energy per
correctly received packet especially for relatively long sessions such
as multimedia ones compared to traditional scalar data ones. This is
mainly due to the fact that a small number of non-interfering paths
allows for better performances than a large number of interfering
ones. The paper is organized as follows. Our incremental
non-interfering multipath protocol is described in section
\ref{proposal}. Some simulation results are shown in section
\ref{simulation}. Section \ref{conclusion} concludes and summarizes
some of our future work.

\section{Maximally Radio-Disjoint\\ Multipath Routing (MR2)}\label{proposal}


Our proposal is an incremental on-demand reactive multipath routing
protocol that makes use of path tables at the sensor nodes. Each
sensor is able to create, maintain and update a path table that
records the different paths to the sink. It contains an entry for each
path with the following fields :

\begin{itemize}
\item {\it pathId}, the path id, corresponds to the last crossed
  sensor in this path from the source to the sink.
\item {\it nextNode}, the next hop toward the sink on this path,
\item {\it metric}, an estimation of the associated quality metric for
this path (hop count, available energy, an estimation of path lifetime
or any other metric depending on the application requirements).
\item {\it inUse}, a flag when set, it indicates that the
  corresponding path is currently in use.
\end{itemize}


A sensor node with respect to the routing process can be either in
active or passive state. In a passive state, as opposed to an active
one, a sensor does not take part in the routing process. From an
energy point of view, a passive node can be put in an idle or even a
sleep mode to save energy and hence increasing the network
lifetime. This depends on the network density and if concurrent
sessions with critical information have to be handled. Clearly, a
tradeoff is to be made between energy saving and serving other
sessions mainly critical ones. We argue that putting some nodes in a
sleep mode is well justified in the case of dense networks as it is
the case of WSN.


\subsection{Route discovery}

In order to build paths, MR2 follows main ideas behind existing
routing algorithms in ad-hoc and sensor networks. The sink floods the
network with a request until the sensor, referred to as the source,
having the requested data is reached. A request contains the following
fields:

\begin{itemize}
\item a request sequence number that gives the rank of the currently
  built path for this session,
\item a path id that corresponds to the first crossed sensor from the
  sink by this request,
\item the last crossed node id, 
\item till this node path quality metric 
\item a flag called {\it isRepair} set when a path has to be repaired
  (the currently built path is to replace a broken one)
\end{itemize}

Initially, all sensor nodes are in the active state and route
discovery is initiated by the sink sending a request using its address
as the {\it path id}. Upon the reception of a request, a sensor node
creates a new path entry if the reported {\it path id} does not appear
in its table. If the request originator is the sink, then the {\it
path\_id} of the entry is this node id ; otherwise, it uses the path
id reported by the current request. The {\it nextNode} field is simply
the last crossed sensor by the request and the {\it inUse} flag is set
to zero. The {\it metric} field is also updated depending on its
nature. If for instance, we choose to use hop count as our metric,
then the reported metric is incremented by one before it is
recorded. If the reported {\it path id} is already stored in the path
table, then the corresponding entry is replaced if the current request
provides a better quality metric ; otherwise, the request is simply
ignored. A request needs to be rebroadcast only if it induces path table
update.

Every time, the source receives a request, it records it in its path
table. When all routes from all of its active neighbors are received,
the source proceeds for the selection of one path with the best metric
value. Immediately, it begins transmitting data on this chosen
path. When a sensor node receives a data packet to be forwarded on a
given path $P$, it sets the {\it inUse} flag and sends (only once per
session) a {\it bePassive} message to its neighbors excepting its next
and previous nodes in $P$. On the reception of a {\it bePassive}, a
node changes its state to passive. When an additional path has to be
built, passive nodes do not react to requests and hence will not
take part in the formation of an additional path. In this way, we
ensure a maximum radio disjointness among built paths, thus improving
the application overall throughput.

As described, MR2 is designed in the context of a request-driven
sensor network model where the sink is the originator of a
request. However, it can be easily extended to the event-driven model
where a sensor detecting a target decide to transmit sensed data to
the sink. In fact, paths can be built using the flooding of the first
data packet transmitted by the source toward the source for instance.

\subsection{Data transmission}

An application data packet header mainly contains a sequence number,
the identity of the source in addition to the path id on which this
packet has to be forwarded. Depending on the application, the
construction of an additional path can be initiated by the sink
immediately after receiving the first data packet or when a given
condition occurs. In the case of video transport for instance, the
sink may require building an additional path when the video quality is
poor. An additional path discovery can also be requested when the sink
detects congestion when it experiences larger inter-packet delays than
expected.

When multiple paths are available, the source has to be able to
partition its traffic on the available paths following a given
strategy depending on the application nature and requirements. If
reliability is of a prime importance, one can choose to transmit
redundantly the same flow on more than one path. In contrary, when the
application is semi-reliable where some losses are tolerated as it is
the case of multimedia and in particular image/video delivery, the
traffic can be divided into multiple flows using layered or multiple
description coding. In the former, there is one base layer and several
enhancement layers and in the latter, flows are of equal importance. 

\subsection{Route maintenance}

Route maintenance is the role of the sink. A broken path is detected
by the sink if it does not receive further data on this path for a
given period of time or when it receives a RERR (route Error)
message. This latter is sent by a sensor when its energy level is
under a given threshold. When a path is reputed to be broken, the sink
initiates a new route discovery phase with the {\it isRepair} field
set. The request, in this case, contains an additional field that records
the id of the broken path. In this way, the source is able to know
which entry it has to remove from its path table.

\section{Simulation Results}\label{simulation}

Our interference-aware routing protocol (MR2) is implemented using
TOSSIM, a bit level simulator for TinyOS platform. We also implemented
the single path approach in addition to a multipath routing protocol
without interference awareness. In all of the implemented schemes, the
quality metric used is path length given as the number of hops from
the source to the sink. In what follows, we will use terms {\it
Single} and {\it HC} to refer to the single path approach and the
multipath scheme without interference awareness.

In our simulations, we considered a square sensor field of size $1000
\times 1000 m^2$ where a given number $N^2$ of static sensor nodes
ranging from 49 to 625 are deployed in a randomized grid. The sink is
located at the upper right corner (coordinates 1000,1000) and a given
number of transmitting sources are selected randomly. Experiments were
performed and averaged over 100 simulations with different topologies
and two different densities corresponding to what we refer to as
sparse and dense modes. In the former, the transmission radio range
for all nodes is set to $1500/N$ and in the latter to $2000/N$ giving
a mean node degree of 8 and 12 respectively.

All the sensors have same processing capability and the energy
dissipation due to processing was neglected in our simulations. For
communication, we adopted the energy model of \cite{EnergyModel}. To
transmit a k-bit message a distance d, the consumed energy is given by
$ E_{Tx}(k,d) = E_{elec} \ k + \epsilon_{amp}\ k\ d^2$. To receive a
k-bit message, a sensor consumes $ E_{Rx}(k) = E_{elec}\ k $ where
$E_{elec}$ is the dissipated energy by the radio to run the
transmitter or the receiver circuitry and $\epsilon_{amp}$ is the
required energy by the transmit amplifier. The passive nodes are put
in the sleep state and assumed to consume $1/100$ the reception power.



\begin{figure}
  \centerline{
    \includegraphics[width=\linewidth,angle=0]{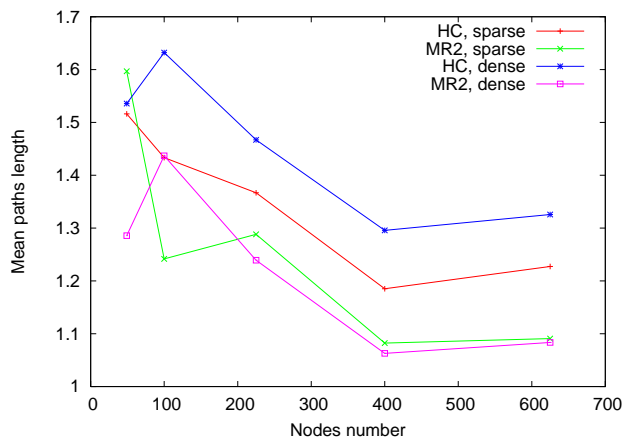}}
  \caption{Paths length ratio with respect to Single}
  \label{Paths-a}
\end{figure}

\begin{figure}
  \centerline{
    \includegraphics[width=\linewidth,angle=0]{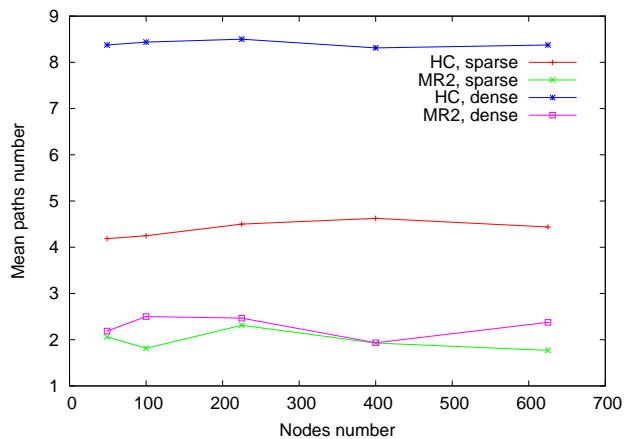}}
  \caption{Number of built paths}
  \label{Paths-b}
\end{figure}

Firstly, we looked at the characteristics of built paths in the
different implemented schemes. Figure \ref{Paths-a} plots the mean
path length (hops number) ratio of built paths in MR2 and HC with
respect to the case of a single path scheme. One fundamental
observation is that MR2 paths are closer to the best path length
(single path scheme) especially when increasing the network
size/density. Figure \ref{Paths-b} gives the mean number of built
paths as a function of the number of nodes. We see that in MR2, we are
able to only build a restricted number of paths (about 2, due to
interference-awareness) compared to HC where more than 4 or 8 paths
are built depending on the network density. Despite that, MR2 achieves
better performances in terms of overall achieved throughput, energy
saving and end-to-end delay even with a few number of paths as will be
shown in what follows.

\begin{figure}
  \centerline{
    \includegraphics[width=\linewidth,angle=0]{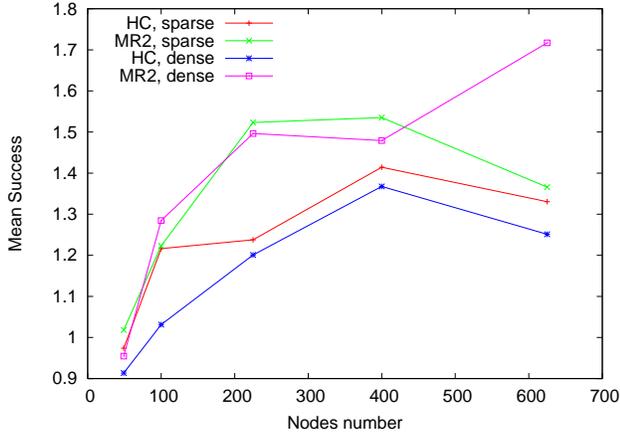}}
  \caption{Success ratio with respect to a single path approach}
  \label{Success}
\end{figure}

\begin{figure}
  \centerline{
    \includegraphics[width=\linewidth,angle=0]{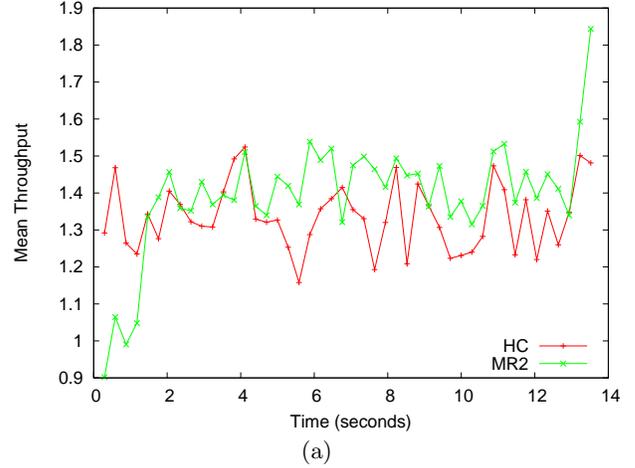}}
  \centerline{(a)}
  \centerline{
    \includegraphics[width=\linewidth,angle=0]{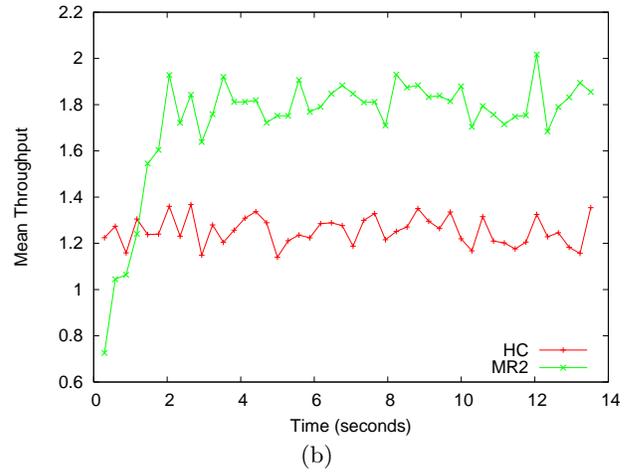}}
  \centerline{(b)}
  
  \caption{Achieved throughput with respect to a single path approach for 625
  nodes (a) sparse mode, (b) dense mode}
  \label{throughput625}
\end{figure}

Figure \ref{Success} plots the overall success ratio of the two
multipath approaches with respect to the single path delivery strategy
as a function of the number of nodes. We can see that MR2 achieves
better success ratio that increases with the network size especially
for highly interfering networks (dense mode). For instance, MR2
improves the success ratio by more than 70\% (with only two paths)
compared to a single path approach instead of only 30\% for HC (with
more than 8 paths). Figure \ref{throughput625} shows the mean
throughput improvement for the different sources as a function of time
and confirms the previous observations. Mainly, we observe that in the
dense mode, MR2 clearly shows better performances. However, in the
sparse mode, achieved throughput is almost the same. Despite that, MR2
still performs better in terms of resources utilization (number of
involved sensors in the delivery process) due to its restricted number
of paths. Table \ref{ActiveNodesNumber} reports the number of involved
nodes in the routing process in the two multipath schemes in both sparse
and dense modes for different network sizes. We can see that in MR2, a
small number of nodes contribute to achieve at least the same level of
performances as HC.

\begin{table}
\caption{Number of involved nodes in the delivery process}
\label{ActiveNodesNumber}

\begin{center}
\begin{tabular}{|c||c|c|c||c|c|c|} 
\hline
Mode & \multicolumn{3}{|c||}{Sparse} & \multicolumn{3}{|c|}{Dense}\\
\hline
Scheme     & HC  & MR2 & ratio & HC & MR2 & ratio\\
\hline
49 nodes  & 24.54  & 12.75 & 0.52 & 24.54 & 9.81 & 0.4\\
\hline
100 nodes & 45.69 & 16.85 & 0.37 & 45.69 & 13.75 & 0.3\\
\hline
225 nodes & 88.02 & 42.6 & 0.48 & 88.02 & 32.33 & 0.37\\
\hline
400 nodes & 130.02 & 49.35 & 0.38 & 130.2 & 35.13 & 0.27\\
\hline
625 nodes & 164.49 & 58.41 & 0.36 & 164.40 & 55.81 & 0.34\\
\hline
\end{tabular}
\end{center}
\end{table}

\begin{figure}
  \centerline{
    \includegraphics[width=\linewidth,angle=0]{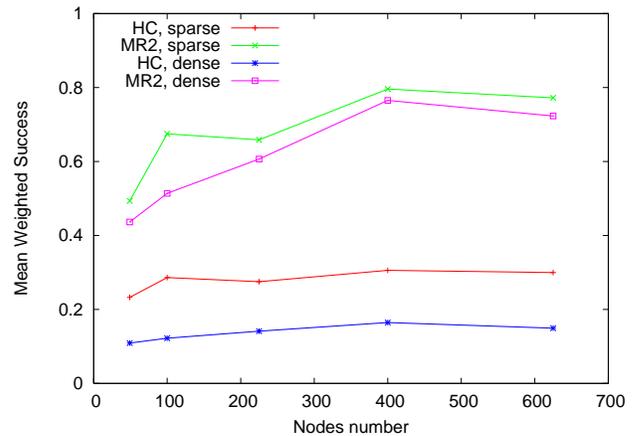}}
  \caption{Success ratio per built path with respect to a single path
  approach.}
  \label{SuccessWR}
\end{figure}

Achieved improvements in MR2 are the result of the absence of
interferences thanks to our incremental approach in building
paths. Figure \ref{SuccessWR} shows how much each built path
contributes in the delivery success. We see that one path in MR2 gets
more than 50\% of success which allows for achieving more throughput
with only two paths than the single path approach. However the
contribution of one path when interferences are frequent (HC) is about
20\%. This observation shows how it is important to build
non-interfering paths.

\begin{figure}
  \centerline{
    \includegraphics[width=\linewidth,angle=0]{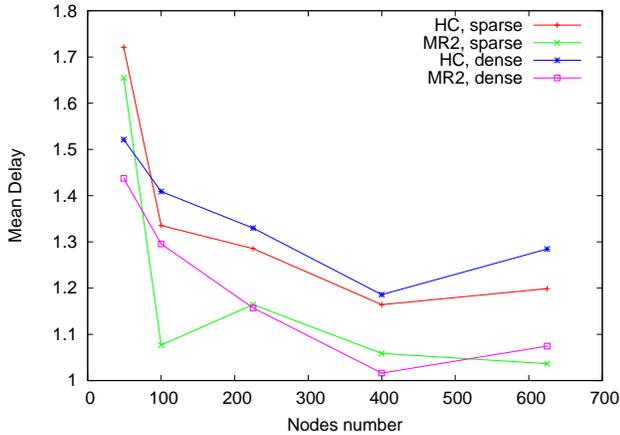}}
  \caption{Delay with respect to a single path approach}
  \label{Delay}
\end{figure}

In terms of the end-to-end delay, figure \ref{Delay} shows that once
again, MR2 performs better since paths length is smaller than those
built by the non-interference aware approach (HC). As already shown
(figure \ref{Paths-a}), MR2 paths length are closer to the smallest
path built in the single path routing. This behavior is more
noticeable when the number of nodes increases. This is a nice feature
for scalability and to avoid big variations in delay between different
paths serving the same session.

\begin{figure}
  \centerline{
    \includegraphics[width=\linewidth,angle=0]{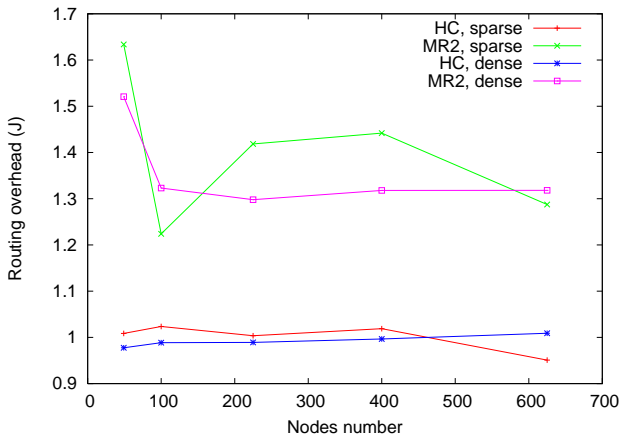}}
  \caption{Overall routing overhead}
  \label{RoutingOverhead-a}
\end{figure}

\begin{figure}
  \centerline{
    \includegraphics[width=\linewidth,angle=0]{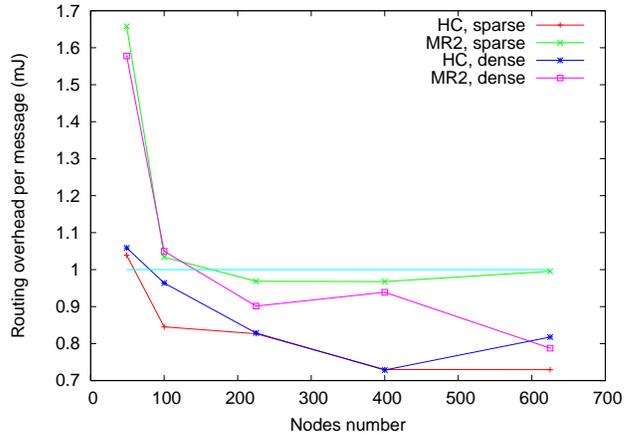}}
  \caption{Per message routing overhead}
 \label{RoutingOverhead-b}
\end{figure}

In order to evaluate the routing overhead introduced by the
incremental approach adopted in MR2, we performed simulations for one
session per source of 15 seconds and others for 5 minutes to see the
energy saving that we can achieve when a session delay increases. We
evaluate routing overhead using the consumed energy by mainly the
request messages. Figures \ref{RoutingOverhead-a} and
\ref{RoutingOverhead-b} plot energy consumption of the routing process
as a function of the network size with respect to single path
routing. As expected, MR2 as shown in figure \ref{RoutingOverhead-a},
introduces about 30\% of additional overhead compared to HC. However,
when we look to the experienced overhead per correctly received packet
(figure \ref{RoutingOverhead-b}), we see that the difference is less
pronounced even if HC introduces less routing overhead.

\begin{figure}
  \centerline{
    \includegraphics[width=\linewidth,angle=0]{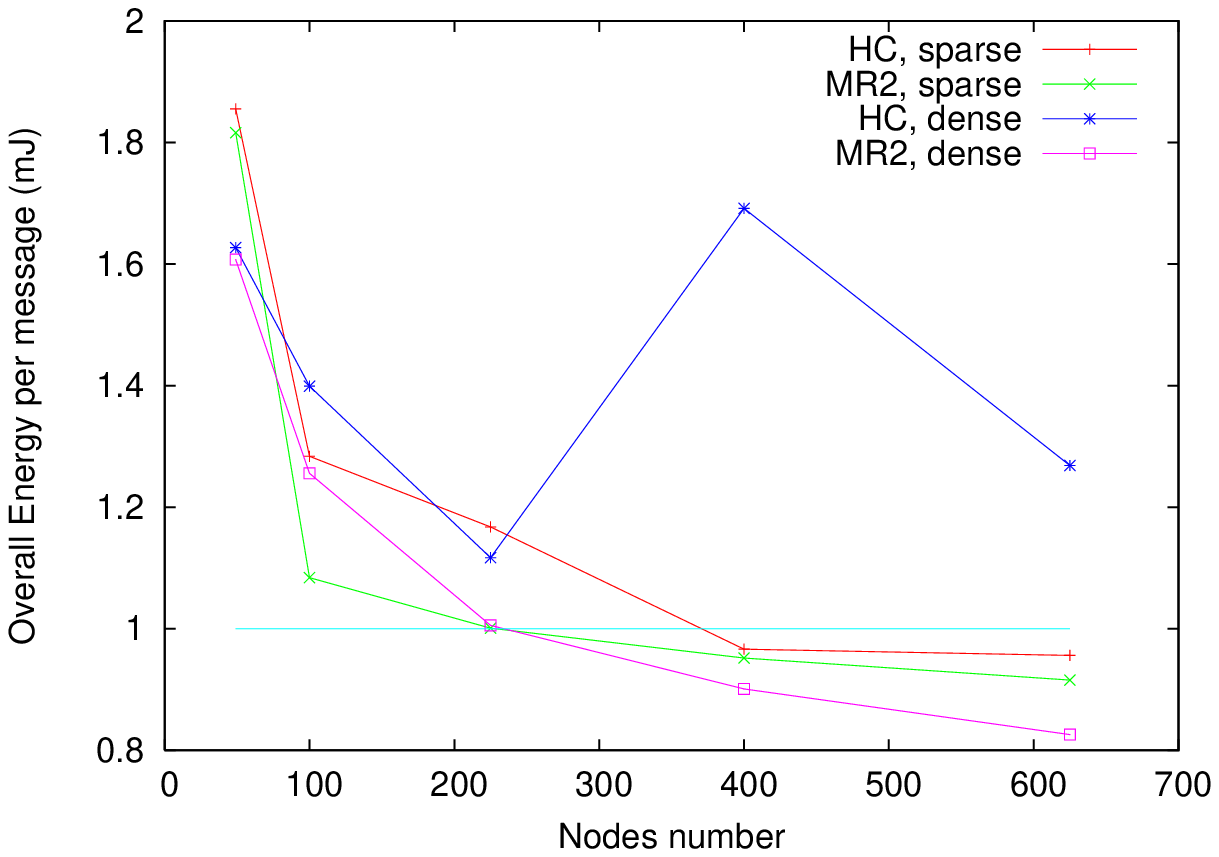}}
  \centerline{(a)}
  \centerline{
    \includegraphics[width=\linewidth,angle=0]{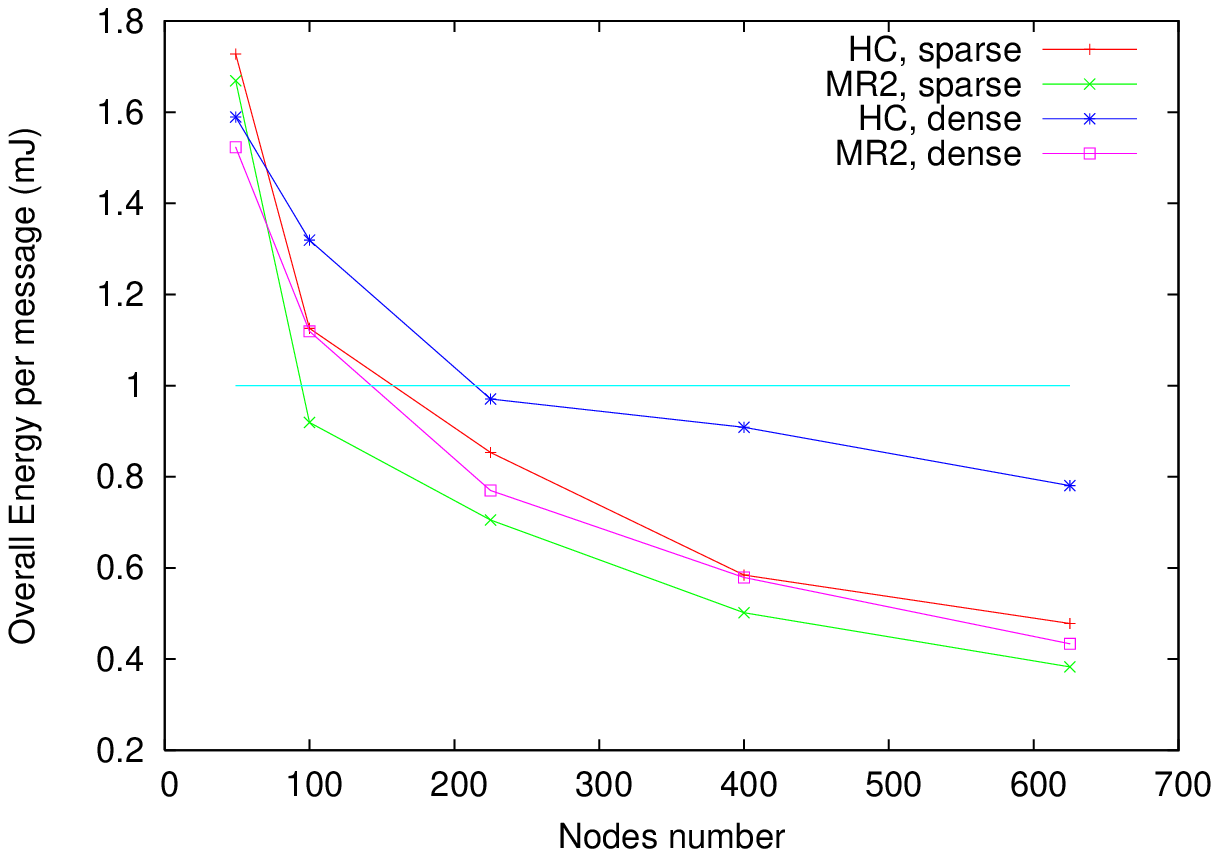}}
  \centerline{(b)}
  
  \caption{Overall consumed Energy per message (including routing
  overhead) in a session of (a) 15 seconds, (b) 5 minutes}
  \label{OverallEperMsg}
\end{figure}

Let us now look to the overall energy consumption per message
including routing overhead. Figure \ref{OverallEperMsg} shows this
amount of energy for the case of short sessions (15 seconds) and long
ones (5 minutes). We can see that for both cases, MR2 consumes less
energy especially for highly interfering and large networks. It is
clear that energy saving will be better if we increase the simulation
time as shown in figure \ref{OverallEperMsg}(b). This is mainly due to
the fact that a given number of nodes (passive ones) are put in the
sleep mode.

\begin{figure}
  \centerline{
    \includegraphics[width=\linewidth,angle=0]{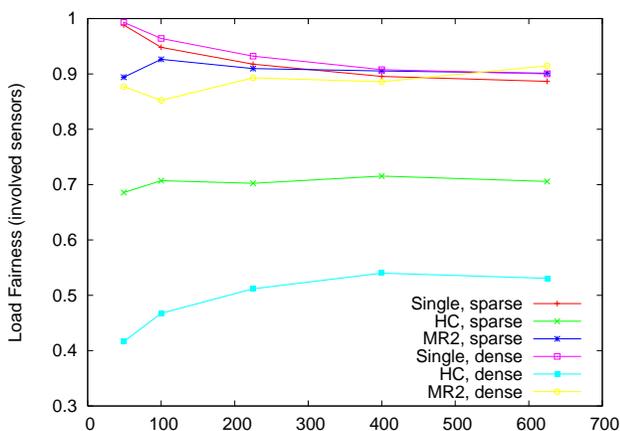}}
  \caption{Load fairness among involved sensors}
  \label{LoadActiveFairness}
\end{figure}

We finally, looked at the load fairness among the sensors involved in
the delivery process in terms of the number of processed messages. We
used $(\sum_{i=1}^{N_a} M_i)^2/(N_a \sum_{i=1}^{N_a} M_i^2)$
where $M_i$ is the number of processed messages by node $i$ and $N_a$
is the number of the involved nodes in delivering data. Figure
\ref{LoadActiveFairness} shows that a better distribution of energy
consumption is achieved in MR2. This allows for avoiding nodes to be
more loaded than others. This decreases the probability that a given
node consumes all of its energy before other ones and consequently, the
network lifetime will be improved.

\section{Conclusion}\label{conclusion}

In this paper, we addressed the problem of interfering paths in the
context of WMSN and considered both intra-session as well as
inter-session interferences. We adopted an incremental approach where
for a given session, only one path is built at once and additional
paths are built when required. Interference awareness and energy
saving are achieved through putting some nodes in a passive
state. Simulation results showed that despite the routing overhead
introduced by the incremental approach we adopted, this was
compensated by the overall achieved throughput and the amount of
consumed energy per correctly received packet especially for
relatively long sessions such as multimedia ones. This is mainly, as
shown, due to the fact that a small number of non-interfering paths
allows for better performances than a large number of interfering
ones. As a future work, we expect to implement both ETX and ETT
metrics in order to compare our approach to those based on these two
metrics. One application to congestion control and image transport is
under development. We also plan to perform experiments in a real WSN
test-bed where our approach will be applied to image/video transport
in order to get more insight into the real benefit we could obtain.

 \section{Acknowledgments}
 This work was supported by French National Research Agenecy (ANR TCAP
 project, No. 06-JCJC-0072).

\scriptsize
\bibliographystyle{abbrv}
\bibliography{wsn}  

\begin{thebibliography}{10}

\bibitem{WMSNSurvey}
I.~F. Akyildiz, T.~Melodia, and K.~R. Chowdhury.
\newblock A survey on wireless multimedia sensor networks.
\newblock {\em Computer Networks}, 51(4):921--960, March 2007.

\bibitem{LQSR}
D.~S.~J. {De Couto}, D.~Aguayo, J.~Bicket, and R.~Morris.
\newblock A high-throughput path metric for multi-hop wireless routing.
\newblock In {\em Proceedings of the 9th {ACM} International Conference on
  Mobile Computing and Networking ({MobiCom} '03)}, San Diego, California,
  September 2003.

\bibitem{MR-LQSR}
R.~Draves, J.~Padhye, and B.~Zill.
\newblock Routing in multi-radio, multi-hop wireless mesh networks.
\newblock In {\em MobiCom '04: Proceedings of the 10th annual international
  conference on Mobile computing and networking}, pages 114--128, New York, NY,
  USA, 2004. ACM.

\bibitem{Highly-DD}
D.~Ganesan, R.~Govindan, S.~Shenker, and D.~Estrin.
\newblock Highly-resilient, energy-efficient multipath routing in wireless
  sensor networks.
\newblock {\em ACM SIGMOBILE Mobile Computing and Communications Review},
  5(4):11--25, 2001.

\bibitem{EnergyModel}
W.~Heinzelman, A.~Chandrakasan, and H.~Balakrishnan.
\newblock Energy-efficient communication protocol for wireless microsensor
  networks.
\newblock In {\em Proceedings of the 33rd Hawaii International Conference on
  System Sciences (HICSS'00)}, January 2000.

\bibitem{LMR}
X.~Hou, D.~Tipper, and J.~Kabara.
\newblock Label-based multipath routing (lmr) in wireless sensor routing.
\newblock In {\em Proceedings of the 6th International Symposium on Advanced
  Radio Technologies (ISART 04)}, Boulder, CO, March 2-4 2004.

\bibitem{RadioDisjoint}
K.~Kuladinithi, M.~Becker, C.~Görg, and S.~Das.
\newblock Radio disjoint multi-path routing in manet.
\newblock In {\em In CEWIT (Center of Excellence in Wireless and Information
  Technology)}, pages 1--2, 2005.

\bibitem{SMR}
S.~Lee and M.~Gerla.
\newblock Split multipath routing with maximally disjoint paths in ad hoc
  networks.
\newblock In {\em IEEE ICC}, volume~10, pages 3201--3205, 2001.

\bibitem{AODV-BR}
S.~J. Lee and M.~Gerla.
\newblock Aodv-br: Backup routing in ad hoc networks.
\newblock In {\em Proceedings of the IEEE Wireless Communications and
  Networking Conference (WCNC 2000)}, Chicago, IL, Sepember 2000.

\bibitem{Yeming2006}
Y.~Lu and V.~W. Wong.
\newblock An energy-efficient multipath routing protocol for wireless sensor
  networks.
\newblock {\em Wiley International Journal of Communication Systems, special
  issue on Energy-efficient Networks Protocols and Algorithms for Wireless
  Sensor Networks}, 2006.

\bibitem{MR-DSR-based}
A.~Nasipuri and S.~Das.
\newblock On-demand multipath routing for mobile ad hoc networks.
\newblock pages 64--70, 1999.

\bibitem{CAM}
I.~Sheriff and E.~M. Belding-Royer.
\newblock Multipath selection in multi-radio mesh networks.
\newblock In {\em BROADNETS}. IEEE, 2006.

\bibitem{SrinivasMobicom03}
A.~Srinivas and E.~Modiano.
\newblock Minimum energy disjoint path routing in wireless ad-hoc networks.
\newblock In {\em Proceedings of the ACM/IEEE International Conference on
  Mobile Computing and Networking (MOBICOM)}, September 2003.

\bibitem{MoorsMeshTech2007}
J.~Tsai and T.~Moors.
\newblock Interference-aware multipath selection for reliable routing in
  wireless mesh networks.
\newblock In {\em MeshTech}. IEEE, 2007.

\bibitem{BoutabaICC06}
S.~Waharte and R.~Boutaba.
\newblock Totally disjoint multipath routing in multihop wireless networks.
\newblock In {\em Proceedings of the IEEE International Conference on
  Communications (ICC 2006)}, Istanbul, Turkey, June 2006.

\end{thebibliography}
\balancecolumns

\end{document}